# Experimental simulation of next-nearest-neighbor Heisenberg chain with photonic crystal waveguide array

F. Qi[1, 2, 3], Y. F. Wang[2, 3], Q. Y. Ma[1, 2, 3, 4], A. Y. Qi[2], P. Xu[5], S. N. Zhu[5], and W. H. Zheng[1, 2, 3, 4]*

[1]State Key Laboratory on Integrated Optoelectronics Lab, Institute of Semiconductors, Chinese Academy of Sciences.

[2]Laboratory of Solid State Optoelectronics Information Technology, Institute of Semiconductors, Chinese Academy of Sciences.

[3]College of Materials Science and Opto-Electronic Technology, University of Chinese Academy of Sciences.

[4]College of Future Technology, University of Chinese Academy of Sciences.

[5]National Laboratory of Solid State Microstructures and College of Physics, Collaborative Innovation Center of Advanced Microstructures, Nanjing University.

**Author Contributes**

F. Qi and Y. F. Wang contribute equally to this work.

**Corresponding Authors**

*E-mail: whzheng@semi.ac.cn.

**Abstract:**

**Next-nearest-neighbor Heisenberg chain plays important roles in solid state physics, such as predicting exotic electric properties of two-dimensional materials or magnetic properties of organic compounds. Direct experimental studies of the many-body electron systems or spin systems associating to these materials are challenging tasks, while optical simulation provides an effective and economical way for immediate observation. Comparing with bulk optics, integrated optics are more of fascinating for steady, large scale and long-time evolution simulations. Photonic crystal is an artificial microstructure material with multiple methods to tune the propagation properties, which are essential for various simulation tasks. Here we report for the**

**first time an experimental simulation of next-nearest-neighbor Heisenberg chain with an integrated optical chip of photonic crystal waveguide array. The use of photonic crystal enhances evanescent field thus allows coupling between next-nearest-neighbor waveguides in such a planar waveguide array, without breaking the weak coupling condition of the coupled mode equation. Particularly, similarities between the model and coherent light propagation could reach 0.99 in numerical simulations and 0.89 in experiment. Localization effect induced by second-order coupling and coupling strengthening with increasing wavelengths were also revealed in both numerical simulations and experiments. The platform proposed here is compatible with mature complementary metal oxide semiconductor technology thus possesses the potential for larger-scale problems and photonic crystals further allows simulations of specific target Hamiltonians.**

Simulating quantum phenomena of inaccessible systems with controllable and observable systems has become an essential method in modern physics[1–4]. Photonic technologies available today are reaching a stage when significant advantages arise for simulating various systems in quantum chemistry, quantum field theory and condensed-matter physics[3,4]. One-dimensional Heisenberg chain with next-nearest-neighbor interaction provides insights into a series of problems, ranging from spin chain model for the ground state properties of many-body systems[5,6] to high-order interaction chain model for triangular lattice materials[6–9]. To date, frustrated Heisenberg spin system has been simulated with bulks optics[5,10], aiming at the ground state properties. A microwave artificial graphene has also been built to study the density of states and the energy shift of Dirac point[9].

Integrated photonic chips enable stable phase control and monolithic integration of large-scale circuits[11–13]. The performance across the chip is robust, repeatable and well understood[14–16]. Simulations of large and complex systems are possible on such platform[17,18]. In particular, silicon photonics is one the most promising platforms due to compact waveguides and mature complementary metal oxide semiconductor (CMOS) technology[19,20]. Sophisticated fabrication control and large circuits for the simulation of long evolution time, large-scale problem are possible on such platform. On the other hand, photonic crystal (PC) is an artificial microstructure material which provides effective methods to tune the propagation properties[21] such as introducing defects and tuning dispersions[22,23]. PCs fabricated on silicon-on-insulator (SOI) wafer are also with high quality. For instance, low loss propagation structure[24] and high Q cavities[25,26] have been realized. A series of photonic quantum information processing schemes have been demonstrated with PCs, including strong interaction of micro-cavity and qubit[27], entangled photon pair generation and engineering[28,29]. Combination of integrated photonics and PC may leads to novel schemes for quantum simulations.

In optics, the simulation can be implemented with either classical light or quantum state of photons due to the bosonic nature of photons[4,30,31]. Particularly, on photonic chips the classical paraxial equation of waveguide is completely analogous to Schrodinger equation[4,32]. As a consequence, the Heisenberg chain (i.e. tight-binding model) can be mapped to weakly coupled waveguide array, where coupling of waveguide modes among waveguides simulates interaction of oscillators and propagation of these modes is equivalent to the evolution of the tight-binding oscillators.

Here we propose a scheme of simulating next-nearest-neighbor (i.e. second-order coupling) Heisenberg chain with an integrated PC waveguide array fabricated on SOI. PC is used to enhance coupling between next-nearest waveguides in such a planar array under the weak coupling condition. In the following, we first present the design of the array with evanescent field overlap and band structure analyses. Secondly, we conduct a series of theoretical calculations to indicate that light propagation in such an array can simulate the second-order coupling Heisenberg chain. Lastly, experimental verification is carried out.

We consider a next-nearest-neighbor Heisenberg chain described by (setting $\hbar=1$ )[5,6,33]

$$\hat{H} = \sum_k \left( \beta_k a_k^\dagger a_k + C_{k,k-1} a_{k-1}^\dagger a_k + C_{k,k+1} a_{k+1}^\dagger a_k + C_{k,k-2} a_{k-2}^\dagger a_k + C_{k,k+2} a_{k+2}^\dagger a_k \right). \tag{1}$$

In optics, $\beta_k$ is the potential of the photon remain in site $k$ and $a_k^\dagger$ is the boson creation operator of site $k$. $c_{k,k\pm1}$ and $c_{k,k\pm2}$ are the first-order and second-order coupling constants, respectively. For a uniform array we have $\beta_k=\beta$, $c_{k,k\pm1}=c_1$ and $c_{k,k\pm2}=c_2$. In this model, $\beta$ just appends a phase factor to the amplitude and will vanish in the final probability[34]. Therefore, we can discard this potential factor out since we just consider probability (or intensity) in the following.

A schematic diagram of the PC waveguide array is presented in Figure 1a. Hexagonal lattice is formed by etching air holes into a silicon slab. The array is then constructed by removing one row of air holes every three rows along the Γ−K direction (the X direction in real space). The lattice constant is $a$=394 nm, the air-hole radius is $0.28a$ and the slab thickness is 340 nm. Coupling between waveguides originates from the overlap of the mode field. Each waveguide is a site and the coupling and propagation of the fundamental waveguide modes can be used to simulate the evolution described by Equation 1[30]. Generally, coupling constant can be expressed by the integral[35,36]

$$C_{j,k} = \frac{k_0^2}{2\beta_j} \frac{\iint_S \delta\varepsilon_r(y,z)\psi_j^*(y,z)\psi_k(y,z)dS}{\iint_S |\psi_k(y,z)|^2 dS}, \quad (2)$$

where $k_0$ is the free space wave vector, $\delta\varepsilon_r(y,z)$ is the perturbation of the dielectric constant profile and $\psi_{j,k}(y,z)$ is the mode amplitudes of individual waveguides.

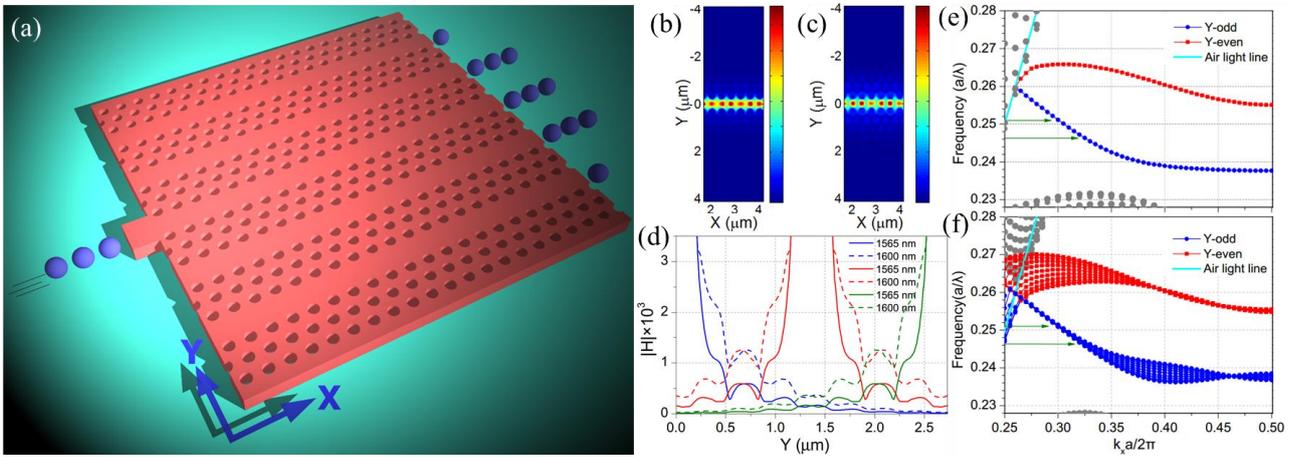

**Figure 1.** (a) Schematic diagram of the second-order coupling PC waveguide array. (b), (c) Modes of the line-defect waveguide at 1565 nm and 1600 nm, respectively. Magnetic field norm is presented here. (d) Illustration of the overlap among three independent line-defect waveguides at two wavelengths as they are separated by three rows of air holes. Their fields are represented by blue, red and olive, respectively. (e), (f) Projected band structures of the PC line-defect waveguide and the PC waveguide array, respectively. Only TE-like modes are considered. The wavelength region marked out by the two olive arrows is 1565 nm–1600 nm when $a$=394 nm.

We present in Figure 1b and c the single line-defect waveguide modes at two wavelengths 1565

nm and 1600 nm, respectively. Corresponding projected band structure is also calculated with 3-dimensional finite-difference time-domain (FDTD) method, as shown in Figure 1e. In the range 1565 nm–1600 nm (marked out by the olive arrows) the modes of the PC waveguide are TE-like, Y-odd, single mode and easily excited by an external strip waveguide[21]. To illustrate the overlap between adjacent and next-nearest waveguides $|H|$ fields of three independent PC line-defect waveguides are plotted in Figure 2d, assuming that the waveguide separation is $2\sqrt{3}a$ (i.e., three rows of air holes between adjacent waveguides). Significant overlap between two next-nearest waveguides indicates that coupling between these waveguides is possible. Furthermore, the overlap is increased at 1600 nm, meaning that coupling is strengthened in long wave region. We then calculated the projected band structure of the waveguide array, as shown in Figure 1f. Two dispersion lines within the band gap (Figure 1e) are broadened to two bands (Figure 1f). The broadening of the Y-odd band is strengthened when wavelength increases, meaning stronger coupling in this region. This is coincidence with the mode overlap analyses.

In longer wave region (>1600 nm), the coupling is so strong that the weak coupling condition of the simple coupled mode equation begins to fail[31,37]. Moreover, there exists a kink point in this region where the symmetry of waveguide modes will change when the Bloch wave vector is scanned through this point[38]. This is interesting for some applications and theoretical investigations[39], but should be avoided here. In shorter wave region (<1565 nm), the broadening of the Y-odd band is minimal, which means that the coupling is very weak and we need a much longer waveguide array to achieve a meaningful evolution time ($zc_1$).

In the following we start to simulate the next-nearest-neighbor Heisenberg chain with classical coherent light propagations. We first calculated light propagation with 3-dimensional FDTD method and extract the intensity distributions at the end of the array. Then these intensity distributions were compared with probability distributions predicted by Equation 1. Generally, we can calculate $c_1$ and $c_2$ using Equation 2. However, for the in-plane one-dimensional waveguide array designed here the adjacent waveguide lies between the next-nearest ones, the decay of the field outside the waveguide core is disturbed[40] and direct calculation using Equation 2 is inaccurate. Another method is comparing the FDTD calculated intensity distributions and probability distributions predicted by Equation 1. Then $c_1$ and $c_2$ can be determined by calculating the similarity[41,42], which is defined by

$$S = \left(\sum_j \sqrt{A_j B_j}\right)^2 / \left(\sum_j A_j \sum_j B_j\right), \qquad (3)$$

where $\{A_j\}$ is the intensity distribution and $\{B_j\}$ is the probability distribution. The maximum of $S$ is 1, meaning that the two distributions are the same. To help with the extraction process, we introduce a ratio $c_2 / c_1 = \delta$.

As a comparison, we also extract coupling constants and similarities when only first-order coupling is considered (i.e., $c_2 = 0$). Two fitting procedures are executed to search for maximum similarity and extract corresponding coupling constants. The first procedure only includes first-order coupling. We scan $c_1$ ($c_2 = 0$) and substitute them to Equation 1 to obtain the probability distributions. Similarities are then calculated and the maximum similarity with corresponding $c_1$ is obtained. The second procedure further includes non-zero $c_2$. In this procedure we simultaneously scan $c_1$ around that obtained in the first procedure and the ratio $\delta$ below 1 (two-dimensional scanning). The similarities are then calculated and the maximum similarity with corresponding $c_1$ and $\delta$ (as well as $c_2$) are extracted.

We calculated light propagations with various exciting wavelengths. In the FDTD calculations, anti-symmetry boundary and symmetry boundary were used in the Y and Z direction, respectively[21]. A strip waveguide, with a width of $\sqrt{3}a$ and a thickness of 340 nm, was connected to the center PC waveguide as the initial exciting port, as presented in Figure 1a. The fundamental TE mode of this waveguide was excited to simulate light propagation in the array. In order to obtain significant ballistic-like propagation[30], evolution time should be reasonably long. Meanwhile, the total waveguide number in the array should be large enough to avoid the reflection or absorption of boundaries[42]. Restricted by our computer's physical memory, a long array with 27 waveguides and a length of $381a$ was used for the short wave range (1565 nm–1585 nm). A short array with 35 waveguides and a length of $233a$ was used for the long wave range (1580 nm–1600 nm).

Maximum similarities at these wavelengths are presented in Figure 2. The overall trend is that similarity decreases when wavelength becomes longer. When second-order coupling is included, the similarities can reach 0.99 in the short wave range (1565 nm–1585 nm). In the long wave range (1585 nm–1600 nm) similarities only with first-order coupling are significantly reduced while similarities with second-order coupling remain high, meaning that second-order coupling is stronger in this region.

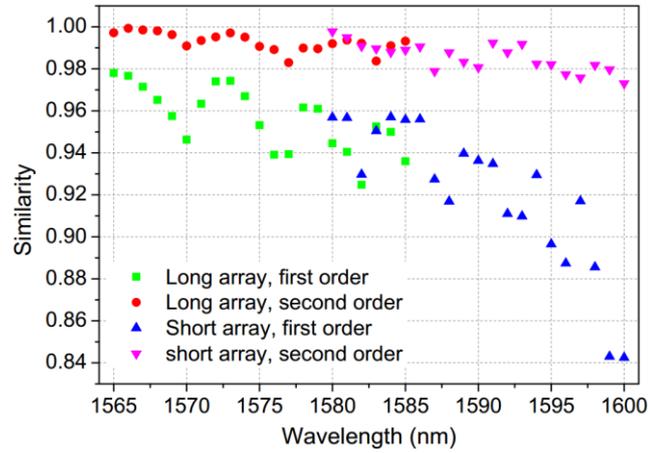

**Figure 2.** Similarities between FDTD calculation and Heisenberg equation. The first-order similarities are obtained in the first fit procedure, and the second-order similarities are obtained in the second fit procedure.

Intensity and probability distributions at three representative wavelengths are also presented in Figure 3. Second-order probability distributions agree well with classical intensity distributions. At short wave region, the coupling is weak thus the evolution time is short; at long wave region, the coupling is strong thus the evolution time is long. As a result, the intensity as well as the probability spread more significant in the long wave region. When only first-order coupling is considered, the coupling constants are $c_1$ =0.015 μm$^{-1}$, 0.020 μm$^{-1}$ and 0.040 μm$^{-1}$ for λ=1565 nm, 1575 nm and 1588nm, respectively. When second-order coupling is included, the first-order coupling constants are $c_1$ =0.014 μm$^{-1}$, 0.020 μm$^{-1}$ and 0.034 μm$^{-1}$ for λ=1565 nm, 1575 nm and 1588 nm, respectively; the ratios are $\delta$ =0.14, 0.16 and 0.18 for λ=1565 nm, 1575 nm and 1588 nm, respectively. Both first-order and second-order coupling constants increase monotonically with increasing wavelength. An important information indicated by Figure 3 is the localization effect: comparing the second-order distribution in the third row (or intensity distribution in the first row) with the first-order distribution in the second row, the probabilities (or intensities) around the center waveguide are significantly enhanced. This effect together with the high similarities between second-order distributions and classical intensity distributions indicate that the designed PC waveguide array is feasible for simulating the next-nearest-neighbor Heisenberg chain.

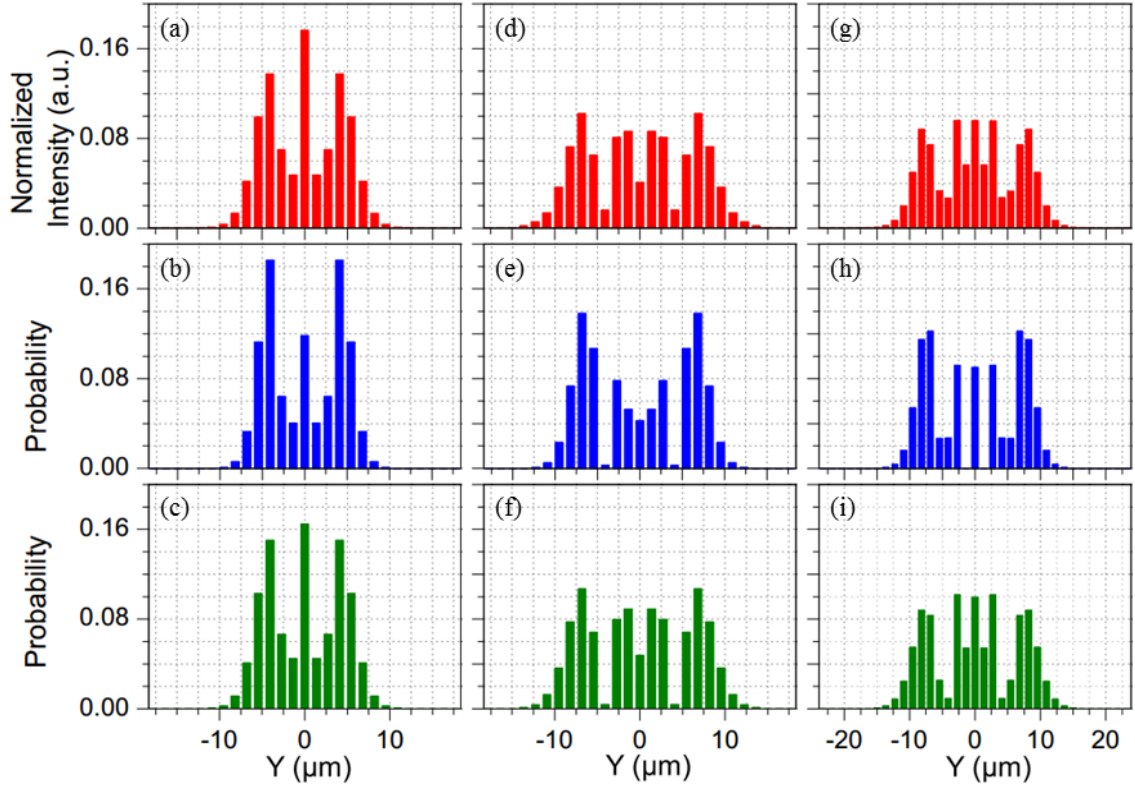

**Figure 3.** Intensity or probability distributions of FDTD calculation and Heisenberg equation. The bar graphs in the first row ((a), (d) and (g)) are FDTD intensity distributions, the bar graphs in the second row ((b), (e) and (h)) are first-order coupling probability distributions. The bar graphs in the third row ((c), (f) and (i)) are second-order coupling probability distributions. The exciting wavelengths are 1565 nm for (a), (b), and (c); 1575 nm for (d), (e) and (f); 1588 nm for (g), (h) and (i).

In conventional waveguide arrays[41,30,43], it is obvious that the coupling will be strengthened when injection wavelength increases. However, in PC waveguide array this coupling strengthening is much more significant. Conventional waveguides are index guiding and the confinement mechanism is total internal reflection. The evanescent field (field outside the waveguide core) is mainly confined to a region nearby side walls, and the increase of the evanescent field amplitude is tiny when wavelength increases (Supporting Figure S2). Furthermore, the waveguide separation should be large enough to fulfill the weak coupling condition. Larger separation also leads to smaller increase in overlap when wavelength becomes longer. While in PC waveguide array field confinement (in the Y direction) is provided by periodically arranged air holes, that is to say, the mechanism of Bragg reflection or photonic band gap[21]. The field outside the PC waveguide core spreads much farther than those of index guiding waveguides. Based on these two factors, the overlap of the evanescent field increases much more when wavelength increases. For instance, $c_1$ obtained in the second fitting procedure are 0.014 μm$^{-1}$ and 0.064 μm$^{-1}$ for $\lambda$=1565 nm and 1600 nm, respectively. As a comparison, $c_1$ of two coupled strip waveguides used in silicon-strip-waveguide-based quantum circuit [20] are 0.0019 μm$^{-1}$ and 0.023 μm$^{-1}$ for λ=1565 nm and 1600 nm, respectively; $c_1$ of two coupled silicon oxynitride waveguides used in a quantum walk waveguide array[41,42,44] are 0.0029 μm$^{-1}$ and 0.0032 μm$^{-1}$ for λ=808 nm and 826 nm, respectively. The coupling constant of the PC waveguide array is strengthened by 4.6 times, while the silicon waveguide's and the silicon oxynitride waveguide's are only strengthened by 1.2 times and 1.1 times, respectively. The details of the calculation is given in detail in the Supporting Information.

The PC waveguide array was fabricated with electron beam lithography (Jeol JBX 6300FS) and inductively coupled plasma etching on the 340 nm-thick top silicon layer of an SOI wafer. The buried oxide layer was 2 μm-thick and the total waveguide number in the array was 37. The substrate was then thinned and diced to expose the facets of the taper and the array. The length of the array was measured to be 180 μm ($457a$) by SEM. As shown in Figure 4(a), a single line-defect waveguide and a tapered strip waveguide were added to the center waveguide of the array as the input port. The line-defect waveguide can reduce the scattering at the intersection between the array and the strip waveguide.

Coherent light from a tunable DFB laser (Santec TSL-510) was end-coupled into the waveguide via a single-mode tapered lens fiber. We used a fiber polarization controller to rotate the polarization to TE and an optical imaging system containing an infrared CCD camera (HAMAMATSU C2741-03) to monitor the intensity distributions at the array's facet, as shown in Figure 4c. The monitored intensity distributions were gray scale image (See Supporting Figure S5), and the intensities of individual waveguides were obtained by integrating the gray scale within corresponding area[30]. As a comparison, we also present the theoretical probability distributions with and without second-order coupling. Due to the high index contrast of SOI and fabrications deviations, the theoretical $c_1$ and $\delta$ predicted above could not be used. We again used the fitting procedure to get these parameters.

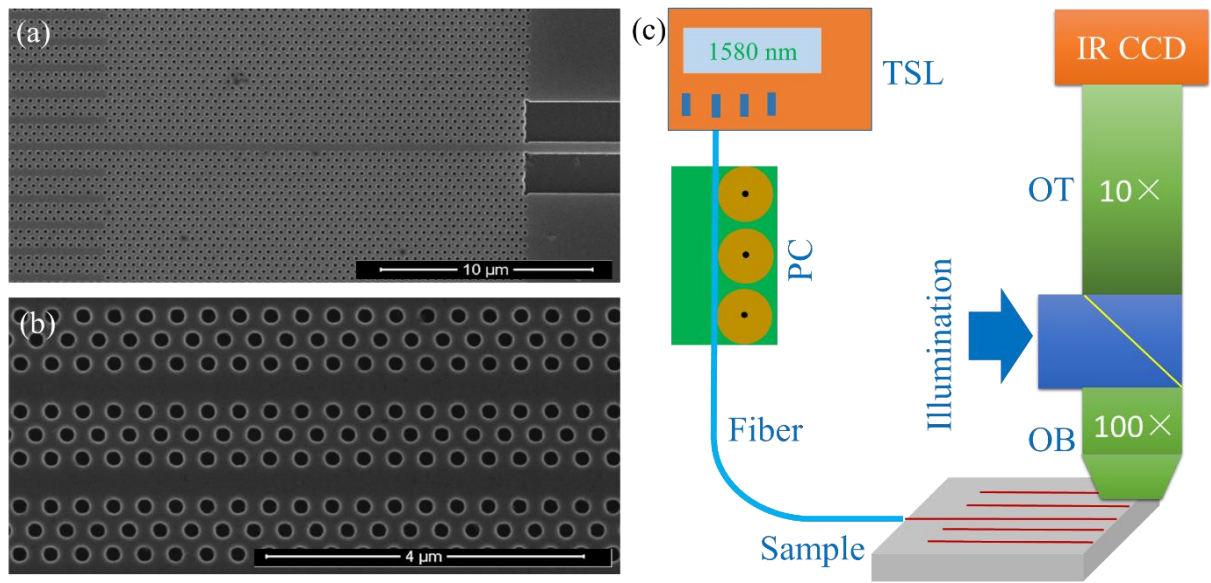

**Figure 4.** (a), (b) Scanning electron microscope (SEM) images of the array. The PC's period is $a=394$ nm, and the radius of the air holes is measured to be 112 nm by SEM. (c) Measurement setup. A single-mode tapered lensed fiber is used to end-coupled light into the waveguide. Since the height of the slab is in sub-wavelength scale, a reasonable amount of light will diffract upwards and collected by the objective right above the end facet of the PC waveguide. TSL: tunable semiconductor laser. PC: polarization controller. OT: optical tube. OB: objective.

As shown in Figure 5, the coupling is strengthened when wavelength increases. Compared with the theoretical distributions only with first-order coupling, measured intensities around the center waveguide is enhanced. Theoretical distributions with second-order coupling predicts this enhancement. In Figure 6 we present the fitted first-order coupling constants and similarities. The ratio $\delta$ obtained in experiment lies between 0.11 and 0.22. As a reference, corresponding values obtained in theoretical analysis above are also presented. In both theoretical analysis and experiment, second-order distributions possess higher similarities with the experimental distributions, reaching 0.89 at 1565 nm. It is straightforward that the obtained coupling constant is slightly increased in the experiment. The reasons are fabricated PCs cannot provide the same strength of light confinement as theoretical prediction, and the slightly increased air-hole radius also leads to stronger coupling. The latter has been verified by our calculation when $r=0.28a$ (Supporting Figure S6). But we should not expect similar coupling strengthening with arbitrary large air-hole radius, since the band gap does not change monotonically with increasing air-hole radius[21]. All in all, we can conclude that coupling enhancement with increasing wavelength and localization effect are observed in experiment.

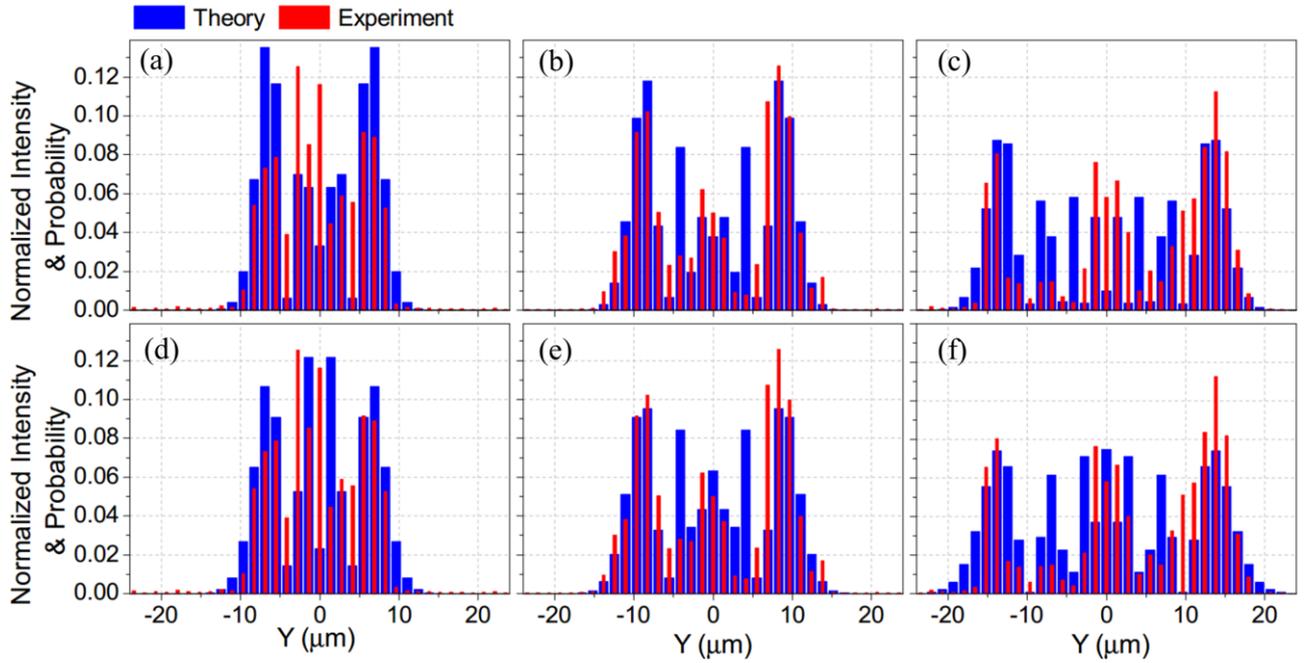

**Figure 5.** Measured intensity distributions at three wavelengths, 1565 nm for (a) and (d), 1575 nm for (b) and (e), 1585 nm for (c) and (f). Corresponding theoretical probability distributions are also presented. In (a), (b) and (c), only first-order coupling is considered. In (d), (e) and (f) second-order coupling is included.

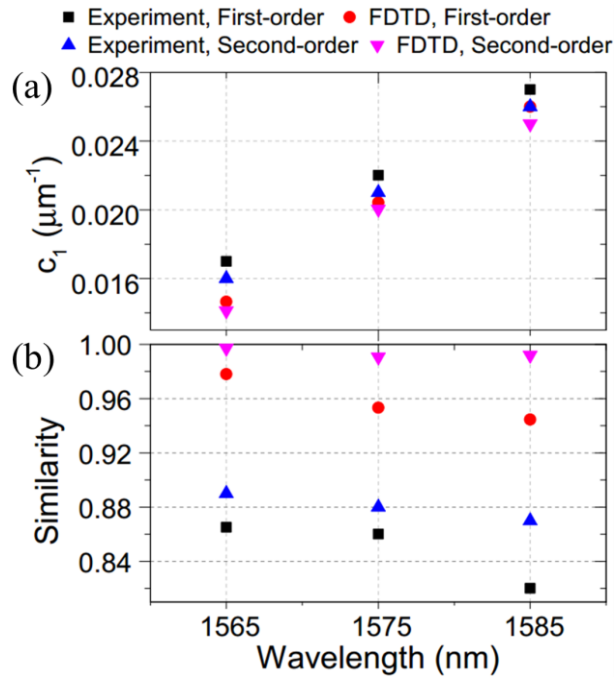

**Figure 6.** First-order coupling constants and similarities. The intensity distributions obtained in the FDTD simulation and experiment were fitted with first-order and second-order coupling Heisenberg chain model, respectively. Corresponding first-order coupling constants and similarities are also presented.

Buried oxide layer of the SOI wafer was not removed in experiment in order to support the large array. This breaks the even symmetry (TE-like) along the Z direction, while theoretical simulations above always assume this symmetry. However, the waveguide layer is thick enough (340 nm) that the field outside the layer is tiny, thus the symmetry breaking is minimal. The dicing process produced a slightly tilted facet at the end of the array, leading to non-uniform waveguide lengths. As a result, the Fabry-Perot interference leads to asymmetric distributions and reduces the similarity. In future works this can be improved by using an etching end facet, instead of the mechanical dicing one.

The coupling strength and the ratio $\delta$ can be tuned by changing the injection wavelengths as demonstrated above. Another way is controlling the air-hole radius since both the size of the band gap as well as the distance between the Y-odd band and the lower edge of the band gap are strongly related to the air-hole radius[21]. Larger gap size or distance means stronger confinement, thus weaker coupling. On the other hand, a much more straightforward way is changing the air-hole rows between adjacent waveguides. With these considerations, we can find points that second-order coupling is small enough that can be ignored, or second-order coupling is much stronger than that involved in this work. However, reduction of the air-hole rows between waveguides needs serious consideration due to the weak coupling condition.

Our results demonstrate that integrated photonic chips based on PC are suitable for the direct observation of second-order-coupling many-body problem and cable of implementing large scale systems with mature CMOS technology. Controllable coupling strengths and $\delta$ with tunable wavelengths will expand the application of a single array, such as simulating electronic properties of single and double layer 2-dimensional materials, or those with different doping levels[8,9,45]. In future works it is also convenient to introduce defects and tune dispersions[46,47] by controlling the position and radius of selected air holes, which will enable optical simulations of open quantum systems and noise enhanced process[48,49], or benefit those where nonlinear process can be included in the simulation tasks[50].

## ACKNOWLEDGEMENTS

This work was supported by the National Key R&D Program of China (Grant No. 2016YFB0402203, 2016YFB0401804, 2016YFA0301102), National Natural Science Foundation of China (Grant

No.61535013, 91321312 and 61137003), Youth Innovation Promotion Association, CAS (Grant No. 2014096). The authors thank Dr. M. Chen's help in EBL and MS. S. Tian's help in ICP etching.## REFERENCES

(1) Trabesinger, A. Quantum Simulation. *Nat. Phys.* **2012**, *8* (4), 263–263.

(2) Georgescu, I. M.; Ashhab, S.; Nori, F. Quantum Simulation. *Rev. Mod. Phys.* **2014**, *86* (1), 153–185.

(3) Aspuru-Guzik, A.; Walther, P. Photonic Quantum Simulators. *Nat. Phys.* **2012**, *8* (4), 285–291.

(4) Longhi, S. Quantum-Optical Analogies Using Photonic Structures. *Laser Photonics Rev.* **2009**, *3* (3), 243–261.

(5) Ma, X.; Dakic, B.; Naylor, W.; Zeilinger, A.; Walther, P. Quantum Simulation of the Wavefunction to Probe Frustrated Heisenberg Spin Systems. *Nat. Phys.* **2011**, *7* (5), 399–405.

(6) Gu, S.-J.; Li, H.; Li, Y.-Q.; Lin, H.-Q. Entanglement of the Heisenberg Chain with the next-Nearest-Neighbor Interaction. *Phys. Rev. A* **2004**, *70* (5), 52302.

(7) Landau, D. P. Critical and Multicritical Behavior in a Triangular-Lattice-Gas Ising Model: Repulsive Nearest-Neighbor and Attractive next-Nearest-Neighbor Coupling. *Phys. Rev. B* **1983**, *27* (9), 5604–5617.

(8) Castro Neto, A. H.; Guinea, F.; Peres, N. M. R.; Novoselov, K. S.; Geim, A. K. The Electronic Properties of Graphene. *Rev. Mod. Phys.* **2009**, *81* (1), 109–162.

(9) Bellec, M.; Kuhl, U.; Montambaux, G.; Mortessagne, F. Tight-Binding Couplings in Microwave Artificial Graphene. *Phys. Rev. B* **2013**, *88* (11), 115437.

(10) Ma, X.; Dakić, B.; Kropatschek, S.; Naylor, W.; Chan, Y.; Gong, Z.; Duan, L.; Zeilinger, A.; Walther, P. Towards Photonic Quantum Simulation of Ground States of Frustrated Heisenberg Spin Systems. *Sci. Rep.* **2014**, *4*, 3583.

(11) Politi, A.; Matthews, J.; Thompson, M. G.; O'Brien, J. L. Integrated Quantum Photonics. *IEEE J. Sel. Top. Quantum Electron.* **2009**, *15* (6), 1673–1684.

(12) Crespi, A.; Ramponi, R.; Osellame, R.; Sansoni, L.; Bongioanni, I.; Sciarrino, F.; Vallone, G.; Mataloni, P. Integrated Photonic Quantum Gates for Polarization Qubits. *Nat. Commun.* **2011**, *2*, 566.

(13) Metcalf, B. J.; Spring, J. B.; Humphreys, P. C.; Thomas-Peter, N.; Barbieri, M.; Kolthammer, W. S.; Jin, X.-M.; Langford, N. K.; Kundys, D.; Gates, J. C.; Smith, B. J.; Smith, P. G. R.; Walmsley, I. A. Quantum Teleportation on a Photonic Chip. *Nat. Photonics* **2014**, *8*, 770–774.

(14) Politi, A.; Cryan, M. J.; Rarity, J. G.; Yu, S.; O'Brien, J. L. Silica-on-Silicon Waveguide Quantum Circuits. *Science* **2008**, *320* (5876), 646–649.

(15) Politi, A.; Matthews, J. C.; O'Brien, J. L. Shor's Quantum Factoring Algorithm on a Photonic Chip.

# Supporting Information: More Numerical Results and Raw Data in Experiment

## 1. Light propagation in PC waveguide array

Here we present two light propagation contours within the wavelength range 1565 nm–1600 nm in Figure S1. This can be regarded as amplitude evolution of one-dimensional next-nearest-neighbor Heisenberg chain. The propagations in Figures S1a and b are similar with the so-called ballistic propagation, which is the key feature of the Heisenberg chain (or continuous time quantum walks) only with first-order coupling[1]. The difference is that here the amplitudes of outermost lobes are not significantly stronger than the amplitude in the middle, which means that the amplitude in the middle is strengthened conversely. This is the result of a small amount of second-order coupling, which generally leads to localization of amplitude[2,3] or probability[4].

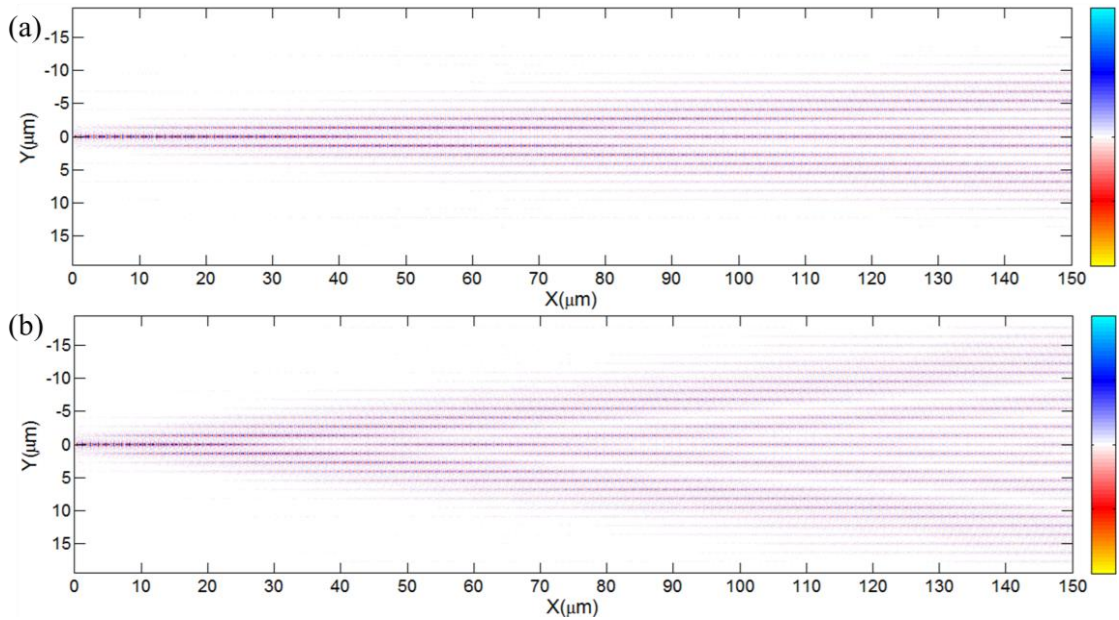

Figure S1. Light propagation in PC waveguide array, calculated by 3-dimension FDTD method. (a) $H_z$ field at 1572 nm. (b) $H_z$ field at 1587 nm.

## 2. Silicon strip waveguide coupler

The waveguide coupler is the same with that used by J.W. Silverstone, et. al.[5] We calculate the modes of the silicon waveguide and fit their amplitude with an exponential function $A \cdot \exp(-B \cdot y)$, as shown in Figure S2. The electric field should be used for this waveguide since the amplitudes of magnetic field on both sides of the wall are tiny. Though the field amplitude outside the silicon core is stronger, it decays quickly. The coefficients are $A=9.9$ and $B=10$ $\mu m^{-1}$ for $\lambda=1565$ nm; $A=9.0$ and $B=9.7$ $\mu m^{-1}$ for $\lambda=1600$ nm. As a comparison, the evanescent field of the PC waveguide is also fitted with the exponential function. The coefficients are $A=0.0022$ and $B=2.0$ $\mu m^{-1}$ for $\lambda=1565$ nm; $A=0.0038$ and $B=1.8$ $\mu m^{-1}$ for $\lambda=1600$ nm. Compared with the PC single line defect waveguide, the evanescent field of the strip waveguide is mainly confined to the nearby waveguide side wall; the evanescent field decays much faster; the amplitude strengthening is also weaker when wavelength increases.

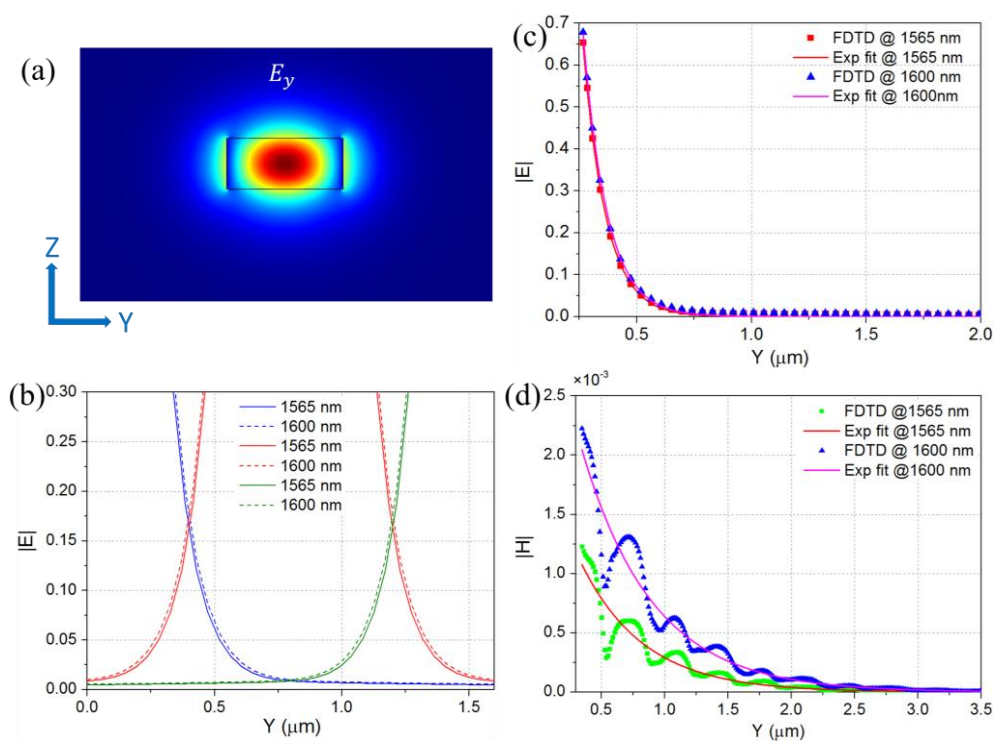

**Figure S2.** (a) The waveguide has a height of 220 nm and a width of 500 nm, embedded in silicon dioxide. The eigenmode at 1565 nm is presented with transverse electric filed, calculated by FEM method. (b) Field overlap between three silicon strip waveguides. The electric field $|E|$ is recalculated by 3D FDTD method to be consistent with the result show in Fig. 2b. (c) The evanescent fields of the strip waveguide and their exponential fitting. (d) The evanescent fields of the PC waveguide and their exponential fitting.

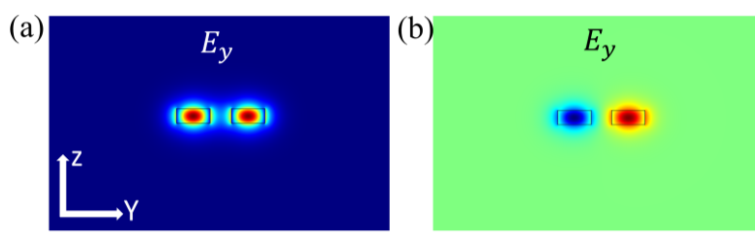

**Figure S3.** Modes of coupled two silicon waveguides. The waveguide dimension is the same with that used in Fig. S1 and the waveguide separation is 800 nm. The modes are calculated by FEM method at 1565 nm.

The overlap of three independent strip waveguides is presented in Figure S2b, assuming that they are separated by 800 nm[5]. The overlap between two adjacent waveguides is significant, while the overlap between two next-nearest waveguides is tiny. The overlap increase is also minimal when wavelength becomes longer. As a result, we should expect a small increase in coupling constant. The coupling constant of two coupled strip waveguide can be directly extracted by computing the symmetric and anti-symmetric modes of the system (Fig. S3). The two modes have different mode effective indexes and the coupling constant is $C = k_0 \left( neff^+ - neff^- \right)/2$ [6]. At λ=1565 nm, the mode effective indexes are 2.4398 and 2.4303 for anti-symmetric and symmetric, respectively. The resulting coupling constant is then 0.019 μm$^{-1}$. At 1600 nm the mode effective indexes are 2.4046 and 2.3930 for anti-symmetric and symmetric, respectively. The resulted coupling constant is then 0.023 μm$^{-1}$. The coupling constant is strengthened by 1.2 times.

## 3. Silicon oxynitride coupler

Another kind of conventional waveguides is weak index confined. A commonly used one is the silicon oxynitride waveguide. Quantum walks of correlated photons have been performed in such a waveguide array[7–9]. A typical waveguide coupler and its modes are shown in Fig. S4[9]. The waveguide core has an index of 1.5217 and the cladding has an index of 1.4532. The mode and field overlap is similar with those of the silicon waveguide, thus we just present the calculated coupling constants in the following. The coupling constant at $\lambda$ =808 nm is 0.0029 μm$^{-1}$. When the wavelength is increased to $\lambda$ =826 nm (1600/1565*808 nm), the coupling constant is 0.0032 μm$^{-1}$. The coupling constant is strengthened by 1.1 times.

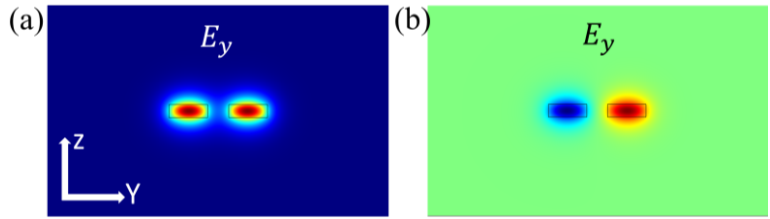

**Figure S4.** Modes of coupled two silicon oxynitride waveguides. The waveguide is 600 nm in height and 1.8 μm in width, and the waveguide separation is 2.8 μm. The modes are calculated by FEM method at 808 nm.

## 4. Experimental Raw data

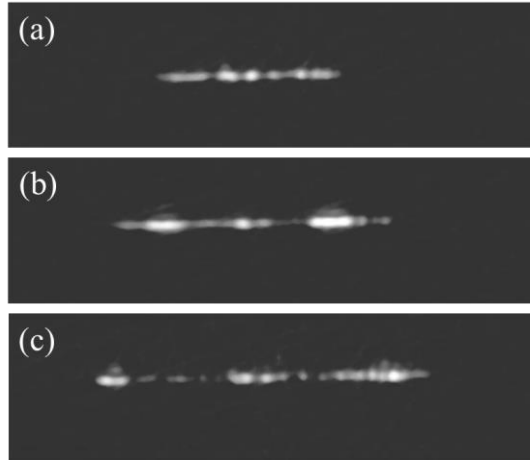

**Figure S5.** CCD measured intensity distributions at the end of the array. The wavelengths are (a) 1565 nm, (b) 1575 nm, (c) 1585 nm.

## 5. Evanescent fields of enlarged air-hole radius

The evanescent field of the PC single line defect waveguide are presented in Fig. S6. When $a$=394 nm, the air-hole radiuses are 110 nm and 112nm. We can find clear evanescent field enhancement when air-holes radius just increased by 2 nm.

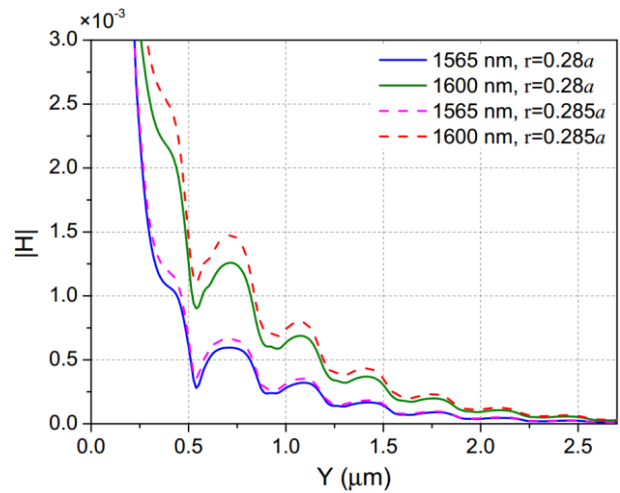

**Figure S6.** Evanescent fields of enlarged air-hole radius.

## 6. Fabrication induced disorder

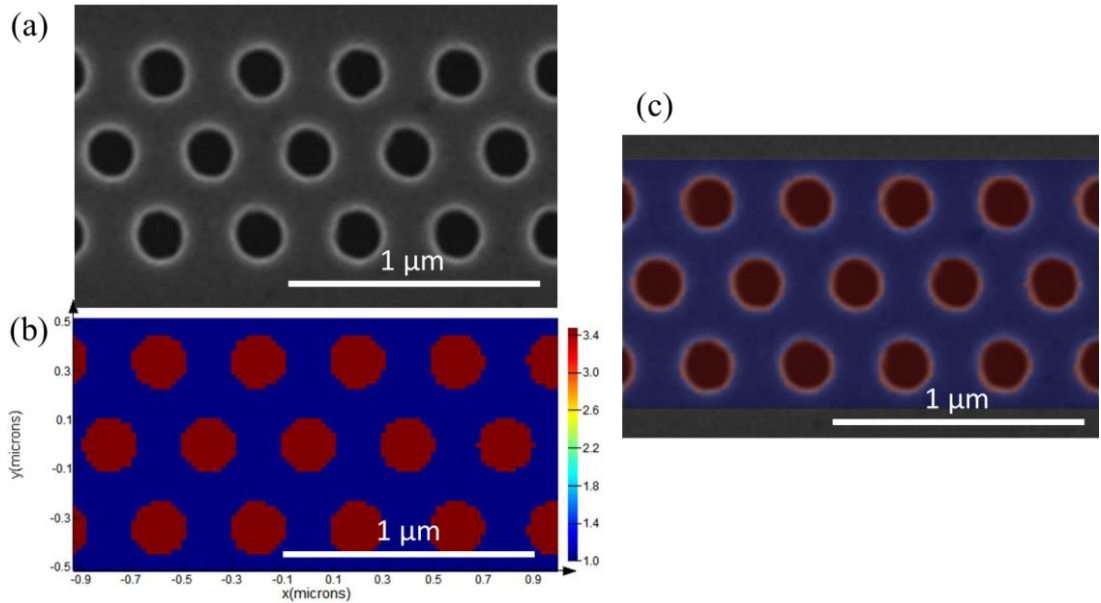

**Figure S7.** Comparison of the fabricated air holes and the air holes used in the 3D FDTD simulation. (a) SEM images of the fabricated air holes. (b) The refractive index of the PC structure used in FDTD simulation, with a grid size of 20 nm. (c) Overlap of (a) and (c).

The fabrication process introduce non-uniformities which also lead to localization. However, the fabrication induced disorder, different from those prior planned, is much weaker[1,10] and has minimal effects in this experiment. This can also be explained by our 3-dimensional FDTD calculation. The grid size is 20nm, which is accurate enough to model light propagation in the PC waveguide array while effectively reflect the deformation of the air holes in fabrication, as shown in Figure S7. The similarities between FDTD simulation and the next-nearest-neighbor Heisenberg chain can reach 0.99.